# Mechanical Control of Individual Superconducting Vortices

Anna Kremen,[†] Shai Wissberg,[†] Noam Haham,[†] Eylon Persky,[†] Yiftach Frenkel,[†] and Beena Kalisky*,[†]

[†]Department of Physics and Institute of Nanotechnology and Advanced Materials, Bar-Ilan University, Ramat-Gan 5290002, Israel

**S** Supporting Information

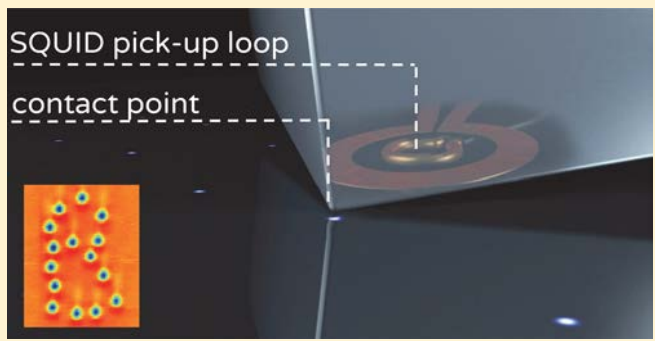

**ABSTRACT:** Manipulating individual vortices in a deterministic way is challenging; ideally, manipulation should be effective, local, and tunable in strength and location. Here, we show that vortices respond to local mechanical stress applied in the vicinity of the vortex. We utilized this interaction to move individual vortices in thin superconducting films via local mechanical contact without magnetic field or current. We used a scanning superconducting quantum interference device to image vortices and to apply local vertical stress with the tip of our sensor. Vortices were attracted to the contact point, relocated, and were stable at their new location. We show that vortices move only after contact and that more effective manipulation is achieved with stronger force and longer contact time. Mechanical manipulation of vortices provides a local view of the interaction between strain and nanomagnetic objects as well as controllable, effective, and reproducible manipulation technique.

**KEYWORDS:** Superconducting vortices, superconductivity, scanning SQUID microscopy, single vortex manipulation

Vortices in superconductors are quantized nanoscale magnetic objects. Motion of individual vortices is controlled by electrical currents (Lorentz force),[1−3] by altering the pinning landscape (e.g., nanostructuring artificial pinning sites[4−6]), or by applying magnetic fields (e.g., magnetic force microscopy[7,8] or scanning superconducting quantum interference device (SQUID) microscopy[9,10]). The effect of mechanical pressure on materials has been the focus of many studies due to its ability to affect and tune electrical and magnetic properties in a variety of systems, including ferroelectric thin films,[11] two-dimensional materials,[12,13] and organic conductors.[14] It was also shown to have a significant effect on superconducting properties such as critical current, critical fields, and critical temperatures.[15] However, the local effect of mechanical stress on vortices in superconductors has not been investigated due to the required locality in both applied stress and view of superconducting properties.

In this work, we used scanning SQUID microscopy to image vortices in thin superconducting films of niobium (Nb) and niobium nitride (NbN) and to map the strength of the superconductor (the diamagnetic response). Here, we demonstrate control over the position of an individual vortex by applying local stress with the tip of our SQUID, without applying current or magnetic field. We determine that vortices are attracted to the contact point and remain stable in their new location. A stable and tunable strain-manipulation technique such as this one can promote applications such as logic elements[16] or spintronic devices[17] and assist the study of vortex dynamics.

Scanning SQUID is a powerful tool for highly sensitive detection of magnetic flux near surfaces. SQUIDs convert flux into measurable electric signal with periodicity of one flux quantum, $\Phi_0$.[18] Magnetic imaging by the SQUID was obtained by recording the magnetic flux as a function of position through a 1−2.6 $\mu$m pickup loop with a sensitivity of 0.7 $\mu\Phi_0$ at 4.2 K.[19] Our SQUID is fabricated on a silicon chip that is polished into a corner. The chip is mounted on a flexible cantilever ($k = 0.35$ N/m) at an angle to the sample, such that the contact point with the sample is at the tip of the SQUID chip (Figure 1a). The size of the contact point is ∼100 nm and varies between chips. By pushing the cantilever into the sample, we apply forces up to 2 $\mu$N, which is well within the mechanically elastic regime of our samples.[20] The capacitance between the cantilever and a static plate determines the contact with the sample and the extent of stress applied. We measured the diamagnetic response from the sample by applying alternating current through an on-chip field coil, located around the pickup loop, and measuring the sample's response to the locally generated magnetic field. When the SQUID chip is brought closer to the surface of a superconducting sample, the superconductor rejects the applied field and the signal measured by the SQUID pickup loop decreases, reflecting the sample's diamagnetism.

Nb and NbN thin films were deposited via direct current magnetron sputtering on silicon substrates with 1 $\mu$m of thermal oxide. Three types of samples were studied: 100 nm thick Nb ($T_C = 8.2$ K), 50 nm thick Nb ($T_C = 7.9$ K), and 30 nm thick NbN ($T_C$ = 11.2 K). Measurements of all samples were made at 4.2 K, unless otherwise noted.

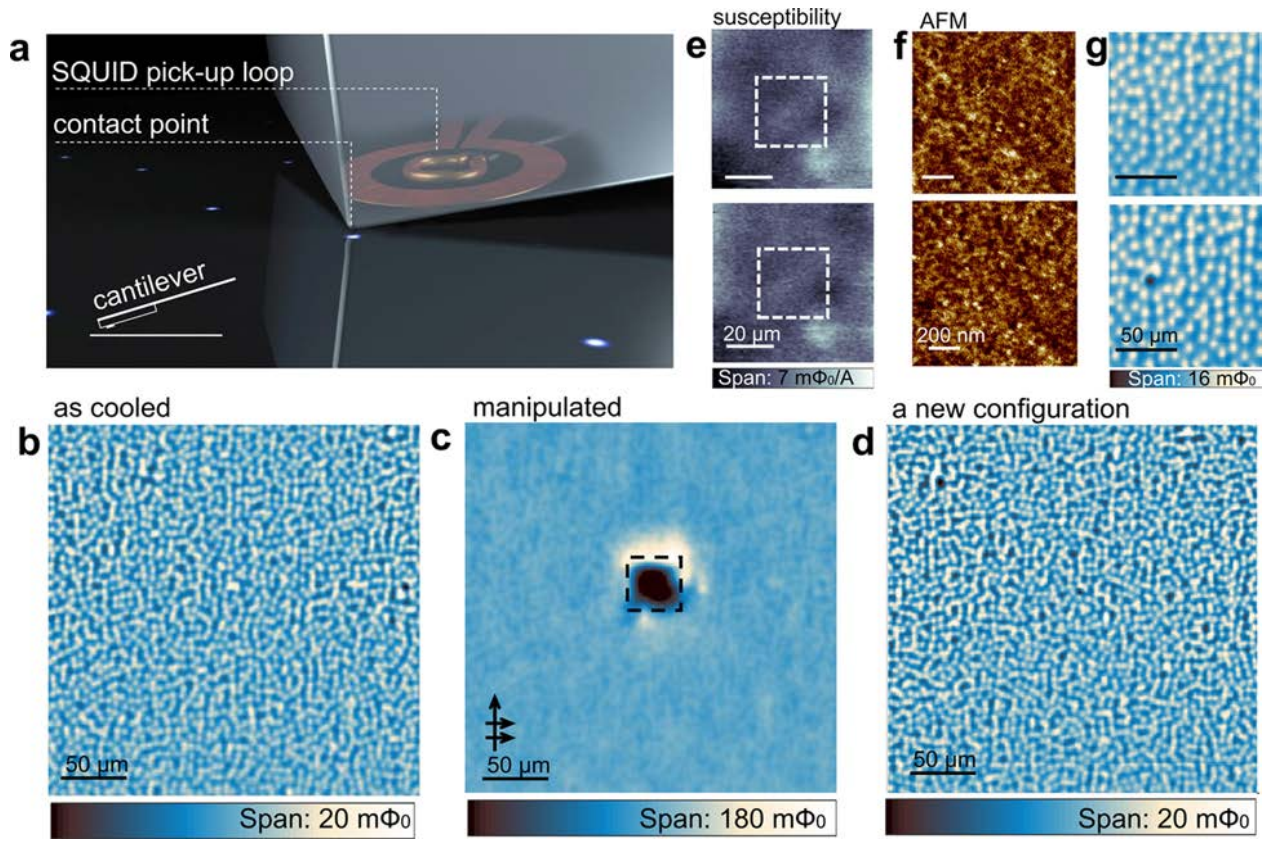

**Figure 1.** Local contact affects vortex configuration in a thin superconducting film. (a) Experimental configuration. Stress is applied with the tip of the SQUID chip by pushing the cantilever (inset) into the sample. (b) Vortex configuration in an NbN thin film imaged at 4.2 K with no contact. The vortices are positive (white), and the flux in each integrates to 1 $\Phi_0$ (Supporting Information). (c) Vortex configuration imaged out of contact after dragging the tip in contact with the sample over a 30 $\mu$m × 30 $\mu$m square (dashed square). The sweep lines in the x-direction were spaced by 250 nm, progressing toward the top of the square (arrows). The strong white signal outside the dark square shows the accumulation of vortices (see Supporting Information for further discussion of the color scale). (d) Renewed vortex configuration on the same area displays no memory of previous manipulations. (e) Diamagnetic response to locally applied field by an on-chip coil, before (top) and after (bottom) the area was scanned in contact in b (dashed square). Susceptibility was measured by applying field of 0.2 G, 800 Hz. The field was not applied during the magnetometry scans in panels b–d. The area was recooled in the presence of 0.1 mG, and no vortices were present during susceptibility measurements. (f) Atomic force microscopy of the area that was scanned in contact (top) and a different area that was not scanned in contact (bottom). We detected no differences in topography or damage to the film. The roughness of the two areas, 0.26 nm root-mean-square, indicates that no damage was made by the contact. (g) Vortex configuration (top), and the same area after we made contact at one point at 8 K, where the SQUID is no longer superconducting but the vortices are still pinned (bottom). Vortices moved toward the contact location (cleared from the darker point). The SQUID was disconnected during contact. Both images were taken with no contact at 4.2 K.

We made contact with the sample by continuously pushing the SQUID tip into the surface while dragging it from one point to another ("sweeping") or briefly pressing it into the surface at a certain point ("tapping"). This physical contact resulted in a movement of vortices to a new location. First, we explored the effect on an ensemble of vortices. We moved vortices out of a small square area by sweeping in contact with the sample; vortices accumulated outside the swept area with almost no vortices left inside (Figure 1b,c). The effect was reversible: a new configuration of vortices, achieved by thermocycling around $T_C$ = 15 K, on the same region of the sample displayed no memory of the contacted region (Figure 1d). This observation implied that the film did not suffer damage during contact, since damaged areas of the film would have been decorated by the new vortex configuration.[21] In addition, local damage to the film results in a locallu reduced diamagnetic response (representing the superfluid density) of the superconducting film. We therefore mapped the landscape of diamagnetic response before and after making contact with a small area on the sample. The spatial modulation on the superfluid density was 1% of the total response of the superconductor (1.2 $\Phi_0$/A), and the landscape did not visibly change as a result of contact (Figure 1e). We further ruled out the possibility of damage to the surface via atomic force microscopy. Comparison of the area that was scanned in contact to an area that was not scanned in contact revealed no changes in topography (Figure 1f). Vortex displacement due to local contact was observed in two samples of Nb and nine samples of NbN and confirmed for thousands of individual, well-separated vortices (Figures 2 and 3). Evidence of vortex motion in contact mode SQUID experiments was previously noticed in weakly pinned materials and was not further investigated.[22,23]

In order to eliminate any magnetic influence from the SQUID, we verified that our results were independent of whether current was flowing in the SQUID and confirmed the same influence of contact on vortices when we contacted the sample at a temperature at which the Nb SQUID was not superconducting but the vortices in NbN were still pinned (Figure 1g). These observations rule out the possibility of influence by proximity of the sample to a superconductor (the Nb SQUID), as well as the possibility of magnetic fields generated by the currents in the device (typically 1−2 mG).

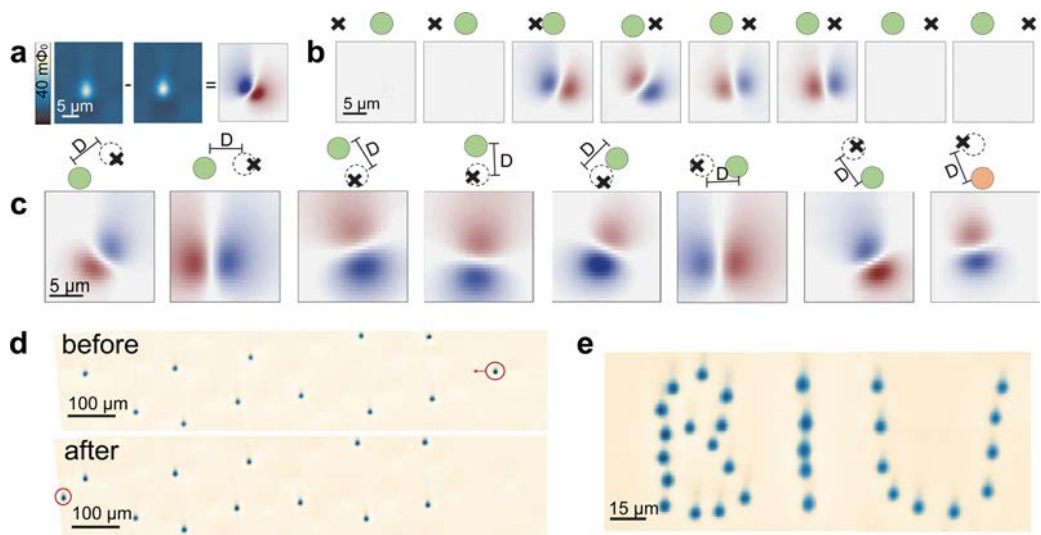

**Figure 2.** A single tap near an isolated vortex attracts the vortex, independent of vortex polarity, enabling excellent control over vortex location. (a) Left and center, two noncontact images of the same vortex before and after tapping the sample once to the left of the vortex. The keyhole shape of an isolated vortex imaged by the SQUID results from the convolution between the magnetic field lines of the vortex and the shape of the SQUID's pick-up loop (see Supporting Information for further discussion). Right, difference between the two images. The vortex moved in the direction of red (old) to blue (new). (b) A series of differential images illustrating the change in vortex location before and after a contact event (tapping). Images were obtained by subtracting consecutive scans. The cartoon above the images shows the position of the contact points (X) relative to the initial location of the vortex (green circle). Taps were spaced by 300 nm and approached the vortex from the left. In effective taps, the vortex moved toward the contact point. The scanned region was not shifted as the vortex moved. (c) Representative differential images taken from a full tapping sequence. Tapping was carried out at various orientations around a vortex within ∼1 $\mu$m. The tapping force was 1 $\mu$N. The location of the contact point is illustrated above each image (X) with respect to the vortex position prior to the tap (green circle) and after the tap (dashed circle). The scale bar on the illustrations marks the distance and direction of movement where $D$ = 0.76, 1.4, 1.6, 1.2, 1.85, 1.4, 0.95, and 0.8 $\mu$m, respectively. Images were taken from the same vortex except for the last one, which was taken from a different vortex with opposite polarity. Note that for the opposite vortex (the last in the series), red (blue) represents the new (old) location. Vortices move toward the contact point regardless of vortex polarity. (d) Vortex configurations before and after we moved one vortex (circled) by 0.95 mm, while the locations of the other vortices did not change. (e) Vortex configuration after we moved vortices to form letters, showcasing the controllability of the technique.

In order to characterize the nature of the interaction between vortices and the mechanical stress at the contact point, we examined the response of a single vortex to stress (Figure 2). We imaged a vortex without contacting the sample, tapped the sample near the vortex with a force of 0.7 $\mu$N, imaged again without contact, and subtracted the images to determine the displacement of the vortex (Figure 2a). In order to determine whether the vortex was drawn to or repelled from the contact point, we performed a series of tapping events along a straight line that passed though the center of an isolated vortex. We imaged the vortex without contact after each tapping event. At first, the vortex did not change its location as a result of the tapping (Figure 2b) but as the tapping approached the vortex the vortex jumped to a new location toward the contact point, implying attraction. This behavior continued when the tapping progressed to the other side of the vortex (shown in Figure 2b by a sequence of differential images). The vortex followed the contact point for a few more taps after which the contact point was too far from the vortex and it ceased to move (Figure 2b). We conclude that the vortex is drawn toward the contact point and that the effective distance of the tap is <2 $\mu$m.

We verified that the interaction is attractive by tapping at various locations around the vortex. Overall, we repeated the tapping experiment on 21 vortices with 215 individual jump events for different vortex polarities, both on Nb and NbN samples. In all experiments, when the contact point was close enough to the vortex, the vortex moved roughly toward the contact point, independent of vortex polarity (Figure 2c). Note that the direction of the displacement was not entirely deterministic; the new vortex location seemed to be determined by the random pinning landscape in the thin film.[24] Vortex manipulation by this method is very effective, and we found no limit to the distance a vortex can move (up to 1 mm in Figure 2d). The vortex always remained stable at its new position. We confirmed stability up to 5 days (data not shown). In addition to stability, this manipulation technique offers excellent control over vortex movements, enabling the design of various vortex configurations (Figure 2e).

In order to determine the onset of the observed effect, we mapped vortex locations as a function of the tip height above the sample. We detected the surface via capacitive sensing and determined the force applied by measuring the cantilever's deflection. When the chip approached the sample, the capacitive reading remained constant. After contact, the capacitance increased sharply, and the contact point was determined by the sharp change in the capacitance slope (Figure 3a). We performed several scans ("sweeping") at selected heights above and in contact with the sample (Figure 3b). Our main observation is that the vortex location never changed before contact; it only changed at or after contact (see cross sections in Figure 3c). We confirmed this behavior in 50 approach sequences (data shown in Figure 3 for one sequence).

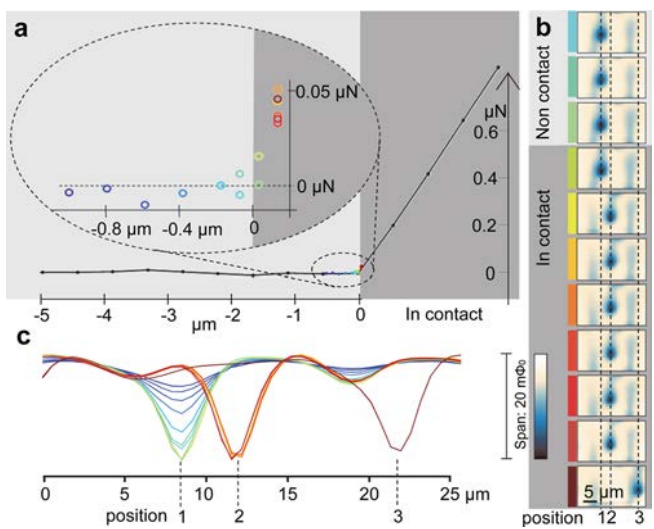

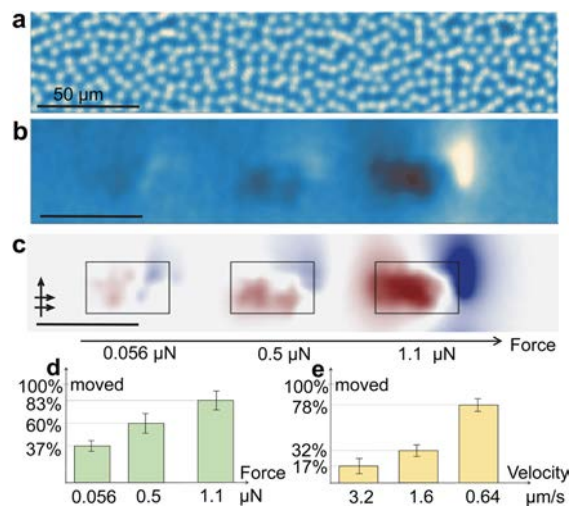

**Figure 3.** Vortices move only after contact with a stress-dependent effectiveness. (a) An approach curve measured via capacitive sensing of the sample plane (black dots). The contact point is reflected by the sharp change in the capacitance slope. (b) Scans at selected heights above and in contact with a single vortex, color coded to correspond to the set of points on the capacitance curve (a). The scan was carried out from left to right and then up to the next row. The vortex moved only after contact (positions 1 to 2) and again after repeated scans in contact (positions 2 to 3). (c) Horizontal cross sections taken at the center of the vortex imaged in b, color-coded to match the relevant points on the capacitance curve. The signal increased during approach, yet the vortex moved only after contact (at maximum signal). The shallow dip observed between positions 2 and 3 is a signal from another vortex (see panel b) that extends in the y-direction due to the keyhole shape of the SQUID (Supporting Information).

**Figure 4.** More vortices move with stronger force and slower scan velocity. (a–c) Noncontact images of vortex configuration before (a) and after (b) contact was made with three subregions sized 36 μm × 17 μm. The scan direction was left to right and then upward (arrows). Each region was swept in contact at a difference force. The vortex moved in the direction of the scan (right and upward). The differential (data in panel b minus the data in panel a) image (c) reveals that more vortices moved at a stronger contact. (d) Stronger force increased the percentage of vortices that moved in each subregion. Vortices moved/total vortices tested: 64/163, 104/165 and 144/173, for 0.056, 0.5, and 1.1 μN, respectively (error bars denote the standard deviation). The experiment was repeated using the same forces, while other experimental parameters remained fixed, such as vortex density and scan line spacing, which strongly influence the number of vortices affected. (e) Slower sweep velocity also increased the percentage of moving vortices. In this experiment, three subareas sized 36 μm × 36 μm were scanned in contact with a constant force of 1 μN. Vortices moved/total vortices tested are 21/121, 38/118, and 88/112 for 3.2, 1.6, and 0.64 μm/s, respectively (error bars denote the standard deviation). The experiment was repeated using the same velocities, while other experimental parameters remained fixed (different values were used in the force experiment in a–d).

We considered the possibility of local heating from friction, because local heating can cause vortex motion.[25] Friction is proportional to both the force and the velocity of the tip in contact; thus, stronger forces or higher sweep velocities are expected to be more effective at moving vortices. To test this prediction, we imaged vortices without contact (Figure 4a) and then swept the tip in contact over three subareas. We applied a different force (0.1, 0.7, and 1.6 μN) to each subarea by pushing the cantilever more strongly into the sample. We imaged the total area again without contact (Figure 4b). Stronger forces moved more vortices in a given area to a new location (Figure 4c,d). Similarly, we tested the effect of sweep velocity by sweeping in contact at a different velocity for each subregion; fewer vortices moved at higher velocities (Figure 4e). Although the strong relationship between applied force and the number of vortices that moved agrees with a friction-based scenario, our observation that faster sweeps displaced fewer vortices contradicts the behavior expected for friction. These results indicate that contact time (longer contact is more effective) is a dominant element in the interaction between the stress and the vortex. We observed this time dependence in both sweeping and tapping experiments. This time dependence eliminates scenarios such as electrostatic discharge by the SQUID chip. We also eliminated the possibility that vortex motion is generated as a result of a temperature difference between the tip and the sample. The SQUID is very sensitive to temperature,[19] but it did not report any temperature change during our experiments.

Local pressure could interact with the vortices via several mechanisms, for example, through a change in the critical temperature. The effect of pressure on $T_C$, $dT_C/dP$, describes the change in the critical temperature $T_C$ as the result of applied pressure $P$. Taking $dT_C/dP$ for a thin film of Nb[26] and the pressure we applied ($\sim 10^8$ Pa) yields a change in $T_C$ of 0.007 K. As a result, the force on a nearby vortex is $<10^{-4}$ pN, which is much lower than the pinning force.[27,28] Furthermore, $dT_C/dP$ is positive in Nb thin films, predicting repulsion rather than the attraction we detected. We also considered the possibility of a small elastic decrease in film thickness due to local pressure. We estimated the thickness variation at the contact point, using the Young's modulus of NbN[20] and Nb,[29] as 0.03% and 0.01%, respectively. The resulting force on a nearby vortex is $10^{-3}$ pN, which is not sufficient to overcome the pinning force. In Nb and NbN films, with a grain size of a few nanometers local stress could also change the spacing between grains and could cause local elastic changes in the pinning landscape or superconducting properties. Both theoretical and experimental efforts are required in order to identify the exact mechanism of the interactions observed here. Scanning SQUID with its sufficient locality, excellent sensitivity, and robustness is an excellent tool for these efforts.

To summarize, we used local physical contact to manipulate individual vortices in a controllable manner over distances up to 1 mm. Scanning SQUID microscopy, which offers locality in both applied stress and view of superconducting properties, revealed that local stress interacts with vortices. Vortices were attracted to the contact point, an effect that became stronger as the applied stress increased and the sweeping velocity decreased. These observations propose a new way to manipulate vortices without magnetic fields or currents and without additional fabrication steps.

## ASSOCIATED CONTENT

**Supporting Information**

The Supporting Information is available free of charge on the ACS Publications website at DOI: 10.1021/acs.nanolett.5b04444.

The SQUID's point spread function, and expanded data presentation for Figures 1 and 2. (PDF)

## AUTHOR INFORMATION

**Corresponding Author**
*E-mail: beena@biu.ac.il.

**Author Contributions**
A.K. and B.K. conceived the investigation. A.K., S.W., E.P., Y.F., and B.K. conducted the SQUID experiments. S.W. obtained the AFM data. A.K., E.P., S.W., N.H., and B.K. analyzed the data and performed calculations. All authors discussed the results and contributed to the manuscript.

**Notes**
The authors declare no competing financial interest.

## ACKNOWLEDGMENTS

We thank A. Sharoni from Bar-Ilan University and N. Katz from the Hebrew University for providing the superconducting films and for helpful discussions. We thank Vladimir Kogan, Alex Gurevich, Eli Zeldov, and Charles Reichhardt for helpful discussions. This research was supported by European Research Council Grant ERC-2014-STG- 639792, Marie Curie Career Integration Grant FP7-PEOPLE-2012-CIG-333799, and Israel Science Foundation Grant ISF-1102/13.